\documentclass{ifacconf}

\usepackage{graphicx}      
\usepackage{natbib}        
\usepackage{epstopdf}
\usepackage{amsmath,amssymb}
\usepackage{colortbl}
\usepackage{xcolor}
\usepackage{tikz}
\usetikzlibrary{positioning, shapes, arrows, calc}
\usetikzlibrary{arrows.meta, positioning}
\usepackage{adjustbox}
\usepackage{pgfplots}
\usepackage[bb=boondox]{mathalfa}
\usepackage{comment}
\usepackage{booktabs}
\usepackage{subfigure}
\newtheorem{assumption}{Assumption}

\newtheorem{remark}{Remark}

\usepackage{pifont}
\newcommand{\xmark}{\ding{55}}
\newcommand{\cmark}{\ding{51}}

\begin{document}
\begin{frontmatter}

\title{Fairness-aware design of nudging policies under stochasticity and prejudices\thanksref{footnoteinfo}} 

\thanks[footnoteinfo]{The work of Mara Tanelli and Camilla Quaresmini was partially supported by MUSA – Multilayered Urban Sustainability Action – project,
funded by the European Union – NextGenerationEU, under the National Recovery and Resilience Plan (NRRP) Mission 4 Component 2
Investment Line 1.5: Strengthening of research structures and creation
of R\&D ‘‘innovation ecosystems’’, set up of ‘‘territorial leaders in R\&D’’
Cod. ECS 00000037.} 

\author[First]{Lisa Piccinin} 
\author[First]{Camilla Quaresmini} 
\author[First]{Edoardo Vitale}
\author[First]{Mara Tanelli}
\author[Second]{Valentina Breschi}

\address[First]{Politecnico di Milano, 20133,
Milan,  Italy (e-mail: \{lisa.piccinin, camilla.quaresmini, mara.tanelli\}@polimi.it).}
\address[Second]{Eindhoven University of Technology, 5600 MB, Eindhoven, The Netherlands (e-mail: v.breschi@tue.nl)}

\begin{abstract}                
We present an injustice-aware innovation-diffusion model extending the Generalized Linear Threshold framework by assigning agents activation thresholds drawn from a Beta distribution to capture the stochastic nature of adoption shaped by inequalities. Because incentive policies themselves can inadvertently amplify these inequalities, building on this model, we design a fair Model Predictive Control (MPC) scheme that incorporates \textit{equality} and \textit{equity} objectives for allocating incentives. Simulations using real mobility-habit data show that injustice reduces overall adoption, while equality smooths incentive distribution and equity reduces disparities in the final outcomes. Thus, incorporating fairness ensures effective diffusion without exacerbating existing social inequities.
\end{abstract}

\begin{keyword} Social networks and opinion dynamics; Systems and Control for Societal Impact;
Fairness in Control; Epistemic fairness; Optimal Control.
\end{keyword}

\end{frontmatter}

\section{Introduction}
The escalating climate crisis, manifesting itself in increasingly extreme weather events and increasing global temperatures and sea levels, demands urgent reductions in greenhouse gas emissions~\citep{aron2022climate}. Green technologies, such as renewable energy sources (e.g., solar photovoltaics), offer a promising pathway toward climate mitigation. However, despite significant public and private investments conceived to make new technologies broadly accessible, their adoption remains limited. In particular, the adoption of low-carbon technologies is still strongly stratified along socio-economic lines, with households with higher incomes, higher education levels, and more stable housing conditions much more likely to invest in \textit{green} technologies than low-income or marginalized households~\citep{burlinson2025socioeconomic}. Indeed, for the latter, adoption is often obstructed by structural barriers such as high upfront technology costs, limited access to credit, and housing precarity.\\ 
Beyond economic aspects, social influence and community dynamics also shape the diffusion of technological (green) innovation. Indeed, individual adoption choices are, for instance, influenced by pressure to conform to peers, deference to high-status individuals, and peer-to-peer information exchange~\citep{van2021editorial, rogers2003diffusion}. Hence, over the years, several opinion dynamics models have been proposed and adapted to describe innovation diffusion. In particular, one can distinguish between averaging models, which represent opinions as continuous variables~\citep{degroot1974reaching, friedkin1999social}, contagion models~\citep{kleinberg2007cascading, kempe2003}, where diffusion is explained as the process of already-infected agents infecting their susceptible neighbors, and binary models, which instead treat individuals' opinions as binary variables~\citep{granovetter1978,holley1975ergodic}. The latter are particularly suited to modeling innovation diffusion, as the final decision to adopt a technology can be naturally represented as a binary variable (i.e., adoption/non-adoption). At the same time, even if these models provide a characterization of the social aspects of innovation diffusion, they often neglect a core feature of the phenomenon and a key consequence of its mutual influence.\\ 
On the one hand, apart from a few exceptions (e.g., see ~\cite{kandhway2014run}), most of these models do not formalize how economic incentives, advertising campaigns, or structural improvements can nudge the diffusion of innovation. Nonetheless, such a characterization is key to enabling a systematic design of strategies to foster the widespread diffusion of innovation. On the other hand, the social aspects of the diffusion process imply that people with lower \textquotedblleft social power\textquotedblright \ may only be marginally involved in the diffusion process, even if they have the knowledge to advise others about the benefits and drawbacks of a specific innovation. At the same time, marginalized individuals may lack information channels through which the benefits of new technologies become visible. This phenomenon, which is known as \textit{epistemic injustice}, i.e., the unfair distribution of credibility due to one's social identity~\citep{fricker2007epistemic}, becomes a form of \textit{testimonial injustice} when it directly concerns the credibility of a speaker being discounted due to their identity (e.g., age, race, or gender) and it is supported by empirical evidence. For instance, women and other marginalized groups are often given less authority even when their expertise is equal~\citep{sieghart2021authority,carli2001gender}, a pattern reproduced and exacerbated in algorithmic systems~\citep{symons2022epistemic,hull2023dirty}. Because such credibility biases shape the people whose opinions are heard and their behavior influences others, they directly affect the social processes through which innovations spread and, thus, cannot be neglected when designing nudging policies.

\emph{Contribution:} Building on the \emph{deterministic} extended Linear Threshold Model (LTM) model~\citep{villa2025epistemic}, which incorporates the epistemic dimension of innovation diffusion as well as the impact of external policies, we propose its stochastic reformulation, drawing on the framework of the Generalized Linear Threshold Model (GLTM)~\citep{mossel2007submodularity}. Accordingly, we still frame adoption as a decision that depends on the fraction (or number) of neighbour peers who have adopted the technology, thereby making this threshold-based mechanism dependent on an activation threshold modeled as a random variable. This formulation allows us to capture the inherent uncertainty in individual decision-making, especially when epistemic factors affect how influence is perceived. We then use the proposed model to design nudging policies that balance the widespread acceptance of the target technology/service, the investment made, and fairness under budget constraints, where fairness is intended as a balance between equality in resource distribution and equity in adoption among population sectors (on the same line as~\cite{villa2025fair}). The presented framework is tested through a simulation study using data from an EU-wide mobility-habits survey~\citep{fiorello2015eu}, and a sensitivity analysis is conducted with respect to fairness objectives.

\emph{Outline:} The proposed epistemic-aware GLTM is presented in Section~\ref{sec:model}, which is then used in Section~\ref{sec:control} to formulate our fair, model predictive policy design problem. Lastly, in Section~\ref{sec:sim_example}, the results of the simulation case study for the proposed strategy are presented and analyzed. The paper concludes with Section~\ref{sec:conclusions}, which summarizes the proposed framework and the obtained results, and provides indications on possible future developments.  

\emph{Notation:} $\mathbb{N}$, $\mathbb{N}_{0}$, and $\mathbb{R}_{+}$ indicate the set of natural numbers, the one of natural numbers including zero, and the set of positive real numbers, respectively. Given any vector $ x \in \mathbb{R}^n $ and matrix $ A \in \mathbb{R}^{m \times n}$, their transposes are indicated as $x^{\top}$ and $A^{\top}$, respectively, while the inverse of $B \in \mathbb{R}^{n \times n}$ is given by $B^{-1}$. A positive (non-negative) definite matrix $A$ is denoted as $A \succ 0$ ($A \succeq 0$). Moreover,
$\|x\|_{B}^{2}=x^{\top} Bx$, 
$x_i \in \mathbb{R}$ denotes the $i$-th component of $x$ and $A_{ij} \in \mathbb{R}$ indicates the element of $A$ in position $(i,j)$. Identity matrices are indicated with $I$. Given a set $\mathcal{A}$, its cardinality is denoted as $|\mathcal{A}|$. Meanwhile, for a random vector $x \in \mathbb{R}^n$, $\mathbb{E}[x]$ indicates its expected value. The logical operator \emph{or} is denoted as $\lor$.  

\section{An epistemic-informed stochastic model of innovation diffusion} \label{sec:model}
Consider a social (influence) network with $N \in\mathbb{N}$ agents, over which we aim to model the diffusion of innovations. Let us formally represent it as a strongly connected graph $\mathcal{G} = (\mathcal{V}, \mathcal{E})$, where $\mathcal{V}$ denotes the set of agents (hence, $|\mathcal{V}|=N$) and $\mathcal{E}$ is the set of edges that encode social connections between them\footnote{$(v,w) \in \mathcal{E}$ if the $v$-th agent is connected to $w$-th one.}. 

Based on the social network $\mathcal{G}$, each agent $v \in \mathcal{V}$ is associated with the set of its neighboring agents 
\begin{equation}
    N_v=\{w \in \mathcal{V}~\mbox{ s.t. } (v,w) \in \mathcal{E}\},    
\end{equation}
which can influence the agent's adoption of the innovation of interest. The latter is mathematically represented by a binary variable $x_v(t) \in \{0,1\}$ which, at each time instant $t \in \mathbb{N}_0$, indicates whether the $v$-th agent has adopted ($x_v(t) = 1$) or not ($x_v(t) = 0$) the technology/service of interest, for all $v \in \mathcal{V}$. In the remainder of the paper, we assume the following about the initial conditions of these binary variables. 
\begin{assumption}[Seed set]
There exists a non-empty set of agents 
\begin{equation}\label{eq:seed_set}
S^*(0) = \{v \in \mathcal{V} | x_v(0) = 1\}\ne \emptyset.    
\end{equation}
This set is denoted as \textit{seed set}, while the agents contained in it are indicated as \textit{initial adopters}. \hfill $\square$
\end{assumption}
In addition to the influence of neighbors, each agent $v \in \mathcal{V}$ has an inherent bias $u^o_v \in [0,1]$ toward the targeted technology. Specifically, the closer $u^o_v$ is to $1$, the more predisposed the $v$-th agent is to embrace the technology or service of interest. Alongside this quantity, we introduce the agent's \emph{reluctance}, which we set at the initial time instant as
\begin{equation}\label{eq:reluctance}
    \rho_v(0) = 1 - u^o_v, ~~~\rho_v(0)\in [0,1],
\end{equation}
and then assume to be shaped by a (eventually personalized) policy $u_v(t) \in [0,1]$ that a stakeholder or policymaker can enact to promote the adoption of a technology/service, for $t \in \mathbb{N}_0$ and $v \in \mathcal{V}$. For instance, as in \cite{villa2024can}, we suppose that the agents' reluctances evolve as
\begin{equation}\label{eq:reluctance_dynamics}
    \rho_v(t+1) = \rho_v(t) + b_v u_v(t),~~~\forall v \in \mathcal{V},~t \in \mathbb{N}_{0},  
\end{equation}
which also depends on a parameter $b_v \in [-1,0]$ that captures the agent's \textit{receptivity} to external interventions, for $v \in \mathcal{V}$. Note that our modeling choice implies that, rather than directly affecting $x_v(t)$, external policies lower individual reluctance to adopt, thereby making the targeted technologies/services more attractive and accessible by depleting socio-economic and practical barriers that prevent adoption in the first place (e.g., economic, environmental, educational, and infrastructural limitations). This choice aligns with actual strategies to promote green services/solutions, e.g., economic incentives to offset high initial costs, awareness campaigns, and dedicated infrastructure. 

The agents' reluctance is also ultimately linked to the innovation diffusion process itself, which we assume to depend on the fraction of neighboring adopters required to convince the $v$-th agent to adopt the new technology in the absence of external incentives, for all $v \in \mathcal{V}$. Indeed, in line with \cite{gao2016general} and \cite{young2009innovation}, we characterize innovation diffusion as an \emph{irreversible cascade} process driven by the relative popularity of the technology within agents' neighbors, i.e.,
\begin{equation}\label{eq:irreversible cascade}
    x_{v}(t\!+\!1)\!=\!\begin{cases}
        1, &\mbox{if } x_{v}(t)\!=\!1\lor\theta_{v}(N_{v}^{\star}(t))\!\geq \!\phi(\rho_{v}(t)),\\ 
        0, &\mbox{otherwise,}
    \end{cases}
\end{equation}
for all $v \in \mathcal{V}$ and $t \in \mathbb{N}_{0}$, where 
\begin{equation}
N^\star_v(t) = N_v \cap S^\star(t), 
\end{equation}
is the set of neighbouring adopters for the $v$-th agent, while 
\begin{equation}
    S^\star(t) = \{ v \in \mathcal{V} | x_v(t) = 1\},
\end{equation}
is the set of overall adopters at instant\footnote{We denote the set of non-adopters at time $t$, i.e., the complementary of $S^\star(t)$, as $\bar{S}^{\star}(t)$.} $t \in \mathbb{N}_0$. Meanwhile, as in \cite{granovetter1978}, $\theta_v$ in \eqref{eq:irreversible cascade} is given by
\begin{equation}\label{eq:mutual_influence}
    \theta_v(N_v^{\star}(t))=\frac{|N_v^{\star}(t)|}{|N_v|} \in [0,1], 
\end{equation}
for all $v \in \mathcal{V}$ and $t \in \mathbb{N}_0$, so that, at each time step, agents synchronously update their adoption state, 
based on the popularity of the technology/service of their neighbors. Therefore, as more people adopt innovation, social pressure to conform increases, eventually leading to more agents adopting it themselves. This mechanism induces a \textit{contagion process} that may propagate throughout the network. This process is also dependent on a reluctance-dependent \emph{thershold} $\phi_v(\rho_{v}(t))$ which, in line with the General Linear Threshold Model (GLTM) proposed in~\cite{gao2016general}, is drawn from a distribution with \emph{finite support} in $[0,1]$, with $\phi_v(\rho_{v}(t))$ being independent of $\phi_w(\rho_{w}(t))$ for all $v,w \in \mathcal{V}$ with $v \neq w$. In this work, we model such thresholds as independent random variables drawn from the Beta distribution~\citep{kagan2024general}, i.e.,
\begin{subequations}\label{eq:threshold2}
\begin{equation}
    \phi_v(\rho_v(t)) \sim \mathrm{Beta}(\alpha_v(t), \beta_v(t);\rho_v(t)),
\end{equation}
whose parameters $\alpha_v(t)=\alpha_v(\rho_v(t))$ and $\beta_v(t)=\beta_v(\rho_v(t))$ depend on the agent’s reluctance $\rho_v(t)$ in \eqref{eq:reluctance}, for each $v \in \mathcal{V}$. Specifically, these parameters are defined as 
\begin{equation}
\alpha_v(\rho_{v}(t)) = \frac{1}{1 - \rho_v(t)}, \quad \beta_v(\rho_v(t)) = \frac{1}{\rho_v(t)}.
\end{equation}
\end{subequations}
Note that this modeling choice implies that node $v$ requires a strong influence from its neighbors before adopting when $\rho_v(t)$ is high, since the Beta distribution concentrates near $1$. Conversely, small values of $\rho_v(t)$ shift the distribution toward $0$, making adoption more likely even under a weaker influence. Therefore, in our setting, the $v$-th agent has a more \emph{conservative} attitude with respect to the technology of interest when $\alpha_v(\rho_v(t)) > \beta_v(\rho_v(t))$, while $\alpha_v(\rho_v(t)) < \beta_v(\rho_v(t))$ reflects a more \emph{progressive} predisposition to adoption, consistent with the distinction between adopter categories discussed in
\citep{rogers2003diffusion}. 
\begin{remark}[On the choice of the Beta distribution]
    As its mean approaches the bounds of the Beta distribution support (i.e., 0 or 1), the variance of $\phi_v(\rho_v(t))$ decreases, $v \in \mathcal{V}$. In turn, this property\footnote{This property has already motivated the application of the Beta distribution in political science contexts, see, e.g.,~\citep{paolino2001maximum}.} implies that the closer  $\mathbb{E}[\phi_v(\rho_v(t))]$ is to 0 (i.e., on average no neighbouring adopters are needed for the $v$-th agent to become an adopter itself) or 1 (namely, all neigbours muts be on average adopters for the $v$-th agent to become an adopter itself), the more individuals' are easily convinced to adopt the technology or firm in their their own preconception against it.  \hfill $\square$
\end{remark}
Moreover, based on our modeling choice, the expected value of the reluctance-based threshold in \eqref{eq:irreversible cascade} is 
\begin{equation}
    \mathbb{E}[\phi_v(\rho_v(t))] = \frac{\alpha_v(\rho_{v}(t))}{\alpha_v(\rho_{v}(t)) + \beta_v(\rho_{v}(t))} = \rho_{v}(t), 
\end{equation}
hence coinciding with the reluctance itself, for all $v \in \mathcal{V}$.
\begin{remark}[Comparison with \cite{villa2024can}]
    This result in expected value is coherent with the thresholds considered in the deterministic cascade model proposed in \cite{villa2024can}, where an agent $v \in \mathcal{V}$ was becoming an adopter whenever 
    \begin{equation} \label{eq:Eu_determistic}
    \frac{|N^*_v(t)|}{|N_v|}\geq \rho_v(t),~~~\forall t \in \mathbb{N}_0,
    \end{equation}
    i.e., when the fraction of its neighbors who are adopters met or exceeded its reluctance. \hfill $\square$
\end{remark}
\begin{remark}[Effect of polices in innovation diffusion]
   Due to our choice of reluctance-based thresholds $\phi_v(\rho_v(t))$, its statistics are linked to the policies $u_v(t) \in [0,1]$, $v \in \mathcal{V}$ and $t \in \mathbb{N}_0$. In particular, since these external measures can only lower individuals' reluctance (see \eqref{eq:reluctance_dynamics} and the properties of $b_v$), they ultimately increase the likelihood that agents will adopt once a progressively smaller fraction of their neighbors has done so, facilitating diffusion. \hfill $\square$
\end{remark}
At the same time, \eqref{eq:irreversible cascade} implies that the adoption is driven by the relative popularity of the technology and that it is assumed to be irreversible, i.e., an adopter remains an adopter as $x_v(\tau) = 1$ for all $\tau > t$ if
$x_v(t) = 1$, for $v \in \mathcal{V}$. 
\begin{remark}[How realistic is irreversibility?]
    Modeling the adoption of an innovation as an irreversible decision may appear unrealistic, since individuals can revise their opinions after experiencing new technologies. Nonetheless, adopting durable technologies (e.g., electric vehicles) often involves substantial upfront investments, making the decision effectively irreversible over the short to medium term\footnote{For example, around 10 years, which corresponds to the average lifespan or payback period of an electric vehicle.}. Hence, reverting this choice within such a timeframe would not only undermine the initial investment but also entail additional costs, making this decision uncommon~\citep{Noel2020}. \hfill $\square$
\end{remark}
\subsection{Modeling the epistemic dimension}
According to \eqref{eq:mutual_influence}, the assumption underlying the formulation in~\eqref{eq:irreversible cascade} is that all individuals have the same ability to shape the adoption of their neighbors. Nonetheless, this modeling choice overlooks the possibility that some agents may have a greater capacity to influence others~\citep{van2011opinion}, both because of stronger social ties (as is the case, for example, with influencers) and their epistemic\footnote{Relating to knowledge.} and social features. Indeed, in reality, not all individuals have the same level of knowledge about the technology of interest, nor are they trusted solely on the basis of that knowledge. This is the case for members of socially marginalized or discriminated groups, who often experience a systematic devaluation of their knowledge due to prejudice~\citep{fricker2007epistemic}. In turn, this implies that these individuals ultimately have less influence on the diffusion of innovations. We thus extend the model in \eqref{eq:irreversible cascade} to account for this phenomenon by changing $\theta_v(N_v^{\star}(t))$ in \eqref{eq:mutual_influence} along the same line of~\cite{villa2025epistemic}. 

To this end, we endow each agent with two additional parameters that characterize its relevance as \textquotedblleft knowers\textquotedblright \ in the diffusion process. Specifically, each agent $v \in \mathcal{V}$ is assigned a \emph{reliability} $\zeta_v \in [0,1]$, representing its actual competence with respect to the new technology and, hence, the weight that its adoption would have on innovation diffusion in the absence of prejudicial distortions. At the same time, in the presence of such (often inevitable) distortions, individuals belonging to \textit{discriminated} groups may be perceived as less credible than their actual competence. Therefore, the perceived competence of an agent becomes its \emph{credibility} $\gamma_v \in [0,1]$, namely the degree to which other agents treat $v$ as epistemically trustworthy. The difference between these two quantities, i.e.,
\begin{equation}
    \Delta_v=\zeta_v-\gamma_v \in [0,\zeta_v],~~\forall v \in \mathcal{V},
\end{equation}
represents what is already introduced in~\cite{villa2025epistemic} as the \emph{credibility deficit}, a form of epistemic injustice arising from socially rooted prejudices. To account for it, we modify \eqref{eq:mutual_influence} as follows: 
\begin{equation}\label{eq:mutual_influence2}
\theta_v (N_v^{\star}(t)) = \frac{1}{|N_v|}\sum_{w \in N_v^{\star}(t)}\!\!\!  (\gamma_w\cdot x_w(t)) \in [0,1],
\end{equation}
thus weighting neighbors' status based on their credibility and, hence, making $\gamma_w$ the key parameter determining the degree to which the $w$-th agent is regarded by its neighbours as a reliable source of information on the technology of interest, with $w \in \mathcal{V}$. This choice enables us to represent the evolution of the individual adoption status as a phenomenon depending on a combination of intrinsic disposition, eventually shaped by external policies, through $\rho_v(t)$, social pressure from connected peers (via $N_v^{\star}(t)$), and the epistemic profile of neighbors due to $\{\gamma_w\}_{w \in N_v^{\star}(t)}$ in \eqref{eq:mutual_influence2}. All the parameters of the proposed cascaded model, including the epistemic dimension (i.e., \eqref{eq:irreversible cascade} with \eqref{eq:mutual_influence2}) are summarized in \tablename{~\ref{tab:models_param}}.
\begin{remark}[Differences with~\cite{villa2025epistemic}]
     The\\ epistemic dimension of innovation diffusion was already introduced in \cite{villa2025epistemic}, with two main differences with respect to the proposed formulation. First, \cite{villa2025epistemic} considers a classical LTM (see~\cite{granovetter1978}), hence restricting to a pure deterministic (and, thus, less realistic) model for the innovation diffusion process. Second, \cite{villa2025epistemic} characterizes the impact of credibility through a credibility-weighted average of neighbors' states, which tends to mask the downweighting of discriminated agents. In contrast, within our stochastic GLTM framework, we explicitly preserve the impact of epistemic biases on influence. \hfill $\square$ 
\end{remark}
\begin{table}[!tb]
  \caption{Parameters and variables in \eqref{eq:irreversible cascade}, for all $v \in \mathcal{V}$ and $t \in \mathbb{N}_0$. \textbf{Symb} indicates their symbol, \textbf{Val} is the interval of parameters' values, while \textbf{Link} highlights their connections.}
    \label{tab:models_param}
    \centering
    \begin{tabular}{llll}
       \hspace*{-.2cm}\textbf{Symb}  \hspace*{-.5cm}& \textbf{Val} \hspace*{-.4cm}& \textbf{Meaning} \hspace*{-.2cm}& \textbf{Link} \\
       \hline
       \hspace*{-.2cm} $x_v(t)$ \hspace*{-.3cm}& $\{0,1\}$ & Status & -\\
        \hline 
       \hspace*{-.2cm} $N_v^{\star}(t)$ & $\{\emptyset,N_{v\!}\}$ & Adopters & Drives $x_v(t)$\\
        \hline 
       \hspace*{-.2cm} $\rho_v(t)$ & $[0,1]$ & Relcutance & Guides $x_v(t)$ based on $N_v^{\star}(t)$\\
        \hline 
       \hspace*{-.2cm} $b_v$ & $[-1,0]$ & Receptivity & Drives $\rho_v(t)$\\
       \hline 
       \hspace*{-.2cm} $u_v(t)$ & $[0,1]$ & Policy & Lowers $\rho_v(t)$ based on $b_v$\\
       \hline 
       \hspace*{-.2cm} $\zeta_v$ & $[0,1]$ & Reliability & - \\
       \hline 
       \hspace*{-.2cm} $\gamma_v$ & $[0,\zeta_v]$ & Credibility & Reduced $\zeta_v$ due to social traits\\
       \hline 
    \end{tabular}
\end{table}

\section{Fair polices to nudge innovation adoption} \label{sec:control}
Building on the model introduced in Section~\ref{sec:model} and inspired by~\cite{villa2025fair}, we now propose a fairness-aware strategy to nudge innovation diffusion adoption rooted in Model Predictive Control (MPC), i.e., to design $\{u_v(t)\}_{v \in \mathcal{V}}$ for $t \in \mathbb{N}_0$. To this end, we assume that policies should be designed without exceeding a finite budget, formalized as follows.   
\begin{assumption}\label{ass:budget}
    The nudging policies $\{u_v(t)\}_{v \in \mathcal{V}}$ at all instants $t \in \mathbb{N}_0$ are designed under a fixed and finite cumulative budget $B \in \mathbb{R}_+$, with $B << \infty$. Hence, the designed policy has to satisfy
    \begin{equation}\label{eq:budget_constr}
        \sum_{v \in \mathcal{V}} u_v(t) \leq B,~~\forall t \in \mathbb{N}_0. \vspace{-.3cm}
    \end{equation} \hfill $\square$
\end{assumption}
This assumption reflects a realistic issue faced by policymakers and stakeholders, \eqref{eq:budget_constr}, but it also implies that the same budget is available at each time step and, hence, that it is not consumed over time. We thus aim to make our hypothesis about the available budget more realistic in the future, assuming it will be depleted over time.

In designing nudging policies, we focus solely on the dynamics of the individual reluctance in \eqref{eq:reluctance_dynamics}, rather than on the stochastic adoption behavior described by \eqref{eq:irreversible cascade}. The motivation for this design choice is two-fold. On the one hand, $\{\rho_v(t)\}_{v \in \mathcal{V}}$ is the only quantity directly affected by the nudging policy, with $t \in \mathbb{N}_0$. On the other hand, such a choice leads to a fully deterministic policy design problem, with stochasticity influencing only the initial design conditions. Indeed, no resource is allocated to agents  adopters, i.e.,
\begin{equation}
    u_v(\tau)=0,~~~\forall \tau\!\geq t~~\mbox{ s.t. } v \in S^{\star}(t)\subseteq \mathcal{V},
\end{equation}
and, hence, the design problem will shrink in size (in terms of the number of decision variables) over time. Accordingly, by pre-fixing a prediction horizon $L\geq 1$, we design the nudging policies by solving the following optimization problem in a receding horizon fashion
\begin{subequations}\label{eq:MPC}
    \begin{align}
    &\underset{\mathcal{R}_{|t},\mathcal{U}_{|t}}{\mathrm{minimize}}~J(\mathcal{R}_{|t},\mathcal{U}_{|t})\label{eq:cost_MPC}\\
        &\qquad ~\mbox{s.t }~\rho_{v|t}(0)=\rho_v(t),\qquad \qquad \qquad~~~ v \!\notin\! S^{\star}(t),\label{eq:initial_condition}\\
        &\qquad \quad~~~\rho_{v|t}(k\!+\!\!1)\!=\!\rho_{v|t}(k)\!+\!b_vu_{v|t}(k),~~v \!\notin\! S^{\star}(t),\\
        & \qquad \quad~~~u_{v|t}(k) \in [0,1],\qquad \qquad \qquad~~~ v \!\notin\! S^{\star}(t),\\
        & \qquad \quad~~~\sum_{v \notin S^{\star}(t)} \!\!u_{v|t}(k)\leq B,\qquad \qquad  k \!\in\! [0,L\!-\!1],
    \end{align}
\end{subequations}
with $\mathcal{R}_{|t}=\{\boldsymbol{\rho}_{|t}(k)\}_{k=0}^{L-1}$ and $\mathcal{U}_{|t}=\{\boldsymbol{u}_{|t}(k)\}_{k=0}^{L-1}$, where $\boldsymbol{\rho}_{|t}(k)$ and $\boldsymbol{u}_{|t}(k)$ stack the reluctances and policies associated to all $v \notin S^{\star}(t) \subseteq \mathcal{V}$ at time $k \in [0,L-1]$, respectively. Therefore, after solving it, we apply only the first nudging policy imposing $u_v(t)=u_{v|t}(0)$ on all $v \notin S^{\star}(t)$ and discard the policies obtained for the next time steps, then repeating the optimization at the subsequent time step with the new initial conditions. 

Note that, according to \eqref{eq:initial_condition}, we assume that policymakers/stakeholders have access to the reluctance $\rho_v(0)$ of each agent $v \notin S^{\star}(0)$ and, therefore, have insights into their inherent bias towards the technology/service of interest. While restricting the practical applicability of the approach, this hypothesis can be relaxed by using data, e.g., population-wide surveys, to estimate these key quantities (see, e.g.,~\cite{villa2024can} for a possible approach). Using such estimates will lead to estimation errors that can affect the policy design scheme, which we systematically address in future work.   

\begin{remark}[How long are nudging policies enacted?]
    As\\ the irreversible cascade proposed in Section~\ref{sec:model} to characterize innovation diffusion is reasonable only in the short/medium periods, we implicitly assume that the policy is enacted over a finite intervention horizon $T\ll \infty$, with $t \in \mathbb{N}_0$. \hfill $\square$
\end{remark}

We are now left to define the loss function $J(\mathcal{R}_{|t},\mathcal{U}_{|t})$ in \eqref{eq:cost_MPC}, which we shape to account for both performance and fairness. Specifically, we impose 
\begin{subequations}\label{eq:MPC_cost_explained}
\begin{equation}
    J(\mathcal{R}_{|t},\mathcal{U}_{|t})\!=\!J^{\mathrm{MPC}}(\mathcal{R}_{|t},\mathcal{U}_{|t})\!+\!J^{\mathrm{FAIR}}(\mathcal{R}_{|t},\mathcal{U}_{|t}),
\end{equation}    
where
\begin{equation}
    J^{\mathrm{MPC}}(\mathcal{R}_{|t},\mathcal{U}_{|t})=\!\!\sum_{k=0}^{L-1}\!\left[\|\boldsymbol{\rho}_{|t}\|_{Q}^{2\!}\!+\!\|\boldsymbol{u}_{|t}\|_{R}^{2}\right]\!+\!\delta V_{\!f}(\boldsymbol{\rho}_{|t}(L)),
\end{equation}
is the classical quadratic MPC cost, including the terminal loss $V_{\!f}(\boldsymbol{\rho}_{|t}(L))$, where $Q \succ 0$ and $R \succ 0$ are positive definite matrices respectively rewarding the annihilation of adoption barriers and savings in deploying the nudging policy. Note that these matrices have time-varying dimensions, as we progressively remove their columns and rows associated with adopters. Meanwhile,  the terminal cost is given by
\begin{equation}\label{eq:terminal_loss}
    V_{\!f}(\boldsymbol{\rho}_{|t}(L))=\|\boldsymbol{\rho}_{|t}(L)\|_{P}^{2},
\end{equation}
where $P \succ 0$ is chosen as the solution of the discrete Riccati equation associated with the reluctance dynamics (see \eqref{eq:reluctance_dynamics}), while $\delta \geq 1$ is a user-selected parameter allowing to have closed-loop stability when using the design policies to foster innovation diffusion (see~\cite[Chapter 2]{rawlings2020model}). Meanwhile, we set 
\begin{equation}
    J^{\mathrm{FAIR}}(\mathcal{R}_{|t},\mathcal{U}_{|t})=J^{\mathrm{equity}}(\mathcal{R}_{|t})\!+\!J^{\mathrm{equality}}(\mathcal{U}_{|t}),
\end{equation}
thus considering also a loss that explicitly accounts for two key facets of fairness in resource allocation, equality (i.e., resources are allocated equally to all) and equity (namely, all agents are given the possibility to achieve the same results). This design choice allows us to consider that $(i)$ pursuing equity might lead to advantaged agents feeling marginalized by the nudging policy, potentially hampering its effectiveness, and that $(ii)$ interventions may unintentionally generate disparities even when the same resources are offered to everyone, since individuals may benefit differently from them depending, e.g., on individual socio-economic status and access to infrastructures. 
\end{subequations}

\subsection{Embedding fairness in nudging policy design}
Pursuing equity means seeking policies that ensure all agents are equally close to adoption at all time instants. Therefore, we characterize $J^{\mathrm{equity}}(\mathcal{R}_{|t})$ as
\begin{subequations}\label{eq:equity}
\begin{equation}
    J^{\mathrm{equity}}(\mathcal{R}_{|t})=\|\boldsymbol{\rho}_{|t}(k) - \bar{\boldsymbol{\rho}}_{|t}(k) \|_{M}^{2},
\end{equation}
where $M\succeq 0$, with $M=0$ indicating that equality is not explicitly considered in policy design, and
\begin{equation}
    \bar{\boldsymbol{\rho}}_{|t}(k)=\left[\frac{1}{|\bar{S}^{\star}(t)|}\sum_{v\notin S^{\star}(t)} {\rho}_{v|t}(k)\right]\mathbb{1},
\end{equation}    
implies that we are steering each non-adopter's reluctance toward the average, thereby mitigating known inequities in innovation diffusion\footnote{It is indeed empirically proven that innovation diffusion tends to increase the social divide.}~\citep{rogers2003diffusion}.
\end{subequations}

On the other hand, striving for equality implies promoting fairness in the distribution of incentives, ensuring that resources are allocated as uniformly as possible across agents. Coherently, the equality loss $J^{\mathrm{equality}}(\mathcal{U}_{|t})$ is defined as 
\begin{equation}
    J^{\text{equality}}(\mathcal{U}_{|t})
    = \| \boldsymbol{u}_{|t}(k) - \bar{\boldsymbol{u}}_{|t}(k) \|^{2}_N,
\end{equation}
where $N \succeq 0$ and 
\begin{equation}
    \bar{\boldsymbol{u}}_{|t}(k)
    = \left[\frac{1}{|\bar{S}^{\star}(t)|}\sum_{v\in S^{\star}(t)} u_{v|t}(k)\right]\mathbb{1}.
\end{equation}
Note that this last term steers the designed nudging policies far from perpetuating distributive inequalities as the innovations spread~\citep{rogers2003diffusion}. As a final note, it is ultimately possible to see that the two terms in $ J^{\mathrm{FAIR}}(\mathcal{R}_{|t},\mathcal{U}_{|t})$ allow us to reveal which strategies lead to the most disparities between agents, both
in the results achieved (e.g., adoption status) and in the nudging resources distributed among individuals.
\begin{remark}[The impact of the epistemic dimension] While the epistemic dimension introduced via \eqref{eq:mutual_influence2} does not explicitly influence the policy design problem in \eqref{eq:MPC}, it does influence $S^{\star}(t)$ for $t\in \mathbb{N}$ and, ultimately, the designed nudging policies.  
\end{remark}

\section{Impact of Nudging Policies: a simulation case study} \label{sec:sim_example}

\begin{figure}[!tb]
    \centering
    \includegraphics[width=0.8\linewidth]{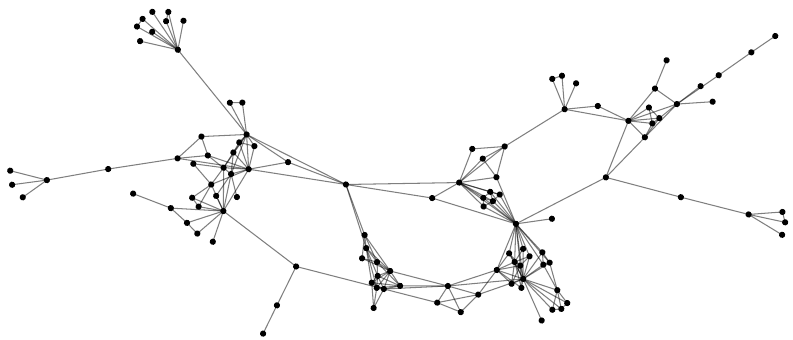}
    \caption{Graph of the considered social network.}
    \label{fig:und_network}
\end{figure}
We now investigate how the nudging policies introduced in Section~\ref{sec:control} affect the evolution of reluctance and the dynamics of adoption of Electric Vehicles (EVs) based on our modeling assumptions, yet using data from a survey on sustainable mobility habits~\citep{fiorello2015eu} to set the parameters of our innovation diffusion model (see \tablename{~\ref{tab:models_param}}). In particular, as in~\cite{villa2025epistemic}, we focus on a subset of $N=112$ respondents whose house is positioned in the metropolitan area of Milan and whose social influence network is estimated from individual mobility habits
and geographical proximity (shown in \figurename{~\ref{fig:und_network} without explicit geographical references). In particular, the initial reluctance $\rho_v(0)$ is estimated based on survey questions targeted to let respondents express their current opinion on EVs. Meanwhile, as shown\footnote{These computations, already presented in \cite{villa2025epistemic}, are intended solely as illustrative examples and should not be considered as definitive guidelines.} in \tablename{~\ref{tab:epistemic_values}} with an example, we use the respondent's educational level (here interpreted as a proxy for perceived competence) to compute the reliability $\zeta_v$ of each agent $v \in \mathcal{V}$, then obtaining $\gamma_v$ by having $\zeta_v$ every time the individual belongs to groups subject to social prejudice based on youth, gender, and income level. While this halving logic is always used to compute individual credibilities, in general, $\zeta_v$ is randomly drawn from a uniform distribution whose bounds are dictated by the agent's education level, for all $v \in \mathcal{V}$. Specifically, such a uniform distribution has support in $[0.7,1]$ if the agent has a high education level, in $[0.4,0.7]$ if the agent has a medium education level, and in $[0,0.4]$ if the agent has a low education level. 

To account for the stochasticity of the adoption process and, in particular, of the relautance-based threshold in \eqref{eq:threshold2}, we design and test the proposed policy design approach over $N_{\mathrm{MC}}=10$ Monte Carlo simulations using the MPC settings in \tablename{~\ref{tab:parameters}} and a simulation horizon of $T=11$ steps. Similarly to \citep{villa2024can}, assuming each step corresponds approximately to 18 days, the spanned time interval in our simulation is around 6 months, which is a sufficiently short time frame in line with our irreversibility assumption.
\begin{table}[tb!]
    \centering
    \caption{Example of a data-driven computation of individuals' epistemic properties based on individuals' Education Level (\textbf{EL}).}
    \label{tab:epistemic_values}
    \begin{tabular}{lccccl}
         & & \multicolumn{3}{c}{\textbf{Discriminated Group}}
         & \\
         \hline
         \textbf{EL} &$\zeta_v$ & \textbf{gender} & \textbf{age} & \textbf{income} & $\gamma_v$\\
         \hline 
         high & 0.8 & \cmark & \xmark & \cmark & $0.8 \cdot 0.5^2 = 0.2$ \\
         \hline
         medium & 0.6 & \xmark & \cmark & \xmark & $0.6 \cdot 0.5^1 =0.3$\\
         \hline 
         low & $0.3$ & \cmark & \cmark & \cmark & $0.3 \cdot 0.5^3=0.0375$\\
         \hline
    \end{tabular}
\end{table}
\begin{table}[!tb]
    \centering
    \caption{Settings of our policy design problem in \eqref{eq:MPC}.}
    \label{tab:parameters}
    \begin{tabular}{ccccc}
         $L$ &   $B$ & $Q$ & $R$ & $\delta$ \\
         \hline
         10 &  50 & $I$ & $I$ & 2 \\
         \hline 
    \end{tabular}
\end{table}
Since the survey does not provide direct information on individuals' receptiveness to external policies, in our simulations we evaluate 4 scenarios. First, we consider an ideal \textbf{{no deficit}} (ND) scenario, where $b_v=-1$ and $\zeta_v=\gamma_v$ for all $v \in \mathcal{V}$. Hence, all agents are fully receptive to nudging policies and are not prejudiced against each other. Then, while still assuming $b_v=-1$, $\forall v \in \mathcal{V}$, we analyze the \textbf{credibility deficit} (CD) case, in which $\gamma_v\leq \zeta_v$ for all $v \in \mathcal{V}$ according to the halving scheme described in \tablename{~\ref{tab:epistemic_values}}. We then restore $\gamma_v= \zeta_v$, yet randomly extract $b_v$ from a uniform distribution within $[-1,0]$, $\forall v \in \mathcal{V}$, thus considering a scenario with a \textbf{receptivity deficit} (RD). Finally, we look at the \textbf{credibility and receptivity deficit} (CRD) case, where $\gamma_v\leq \zeta_v$ and $b_v$ is again randomly extracted from $[-1,0]$ for all $v \in \mathcal{V}$. We further distinguish two cases:
\begin{itemize}
    \item the \textbf{one-sided} design scenario, in which $J^{\mathrm{FAIR}}(\mathcal{R}_{|t},\mathcal{U}_{|t})$ is neglected by setting $M=N=0$;
    \item the \textbf{fair} design scenario, in which $J^{\mathrm{FAIR}}(\mathcal{R}_{|t},\mathcal{U}_{|t})$ is not only accounted for but heavily weighted by setting $M=N=10I$.
\end{itemize}

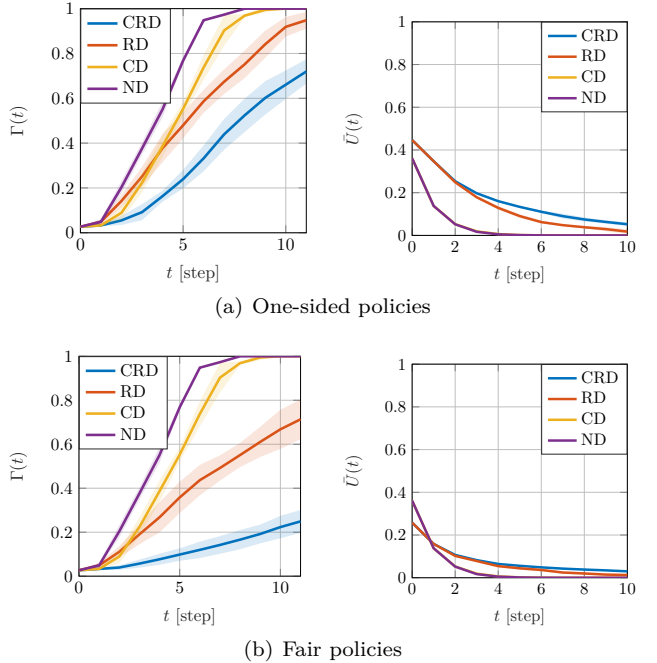
\begin{figure}[!tb]
\centering
\begin{tabular}{c}
 \subfigure[One-sided policies\label{fig:no_fair}]{
 \begin{tabular}{cc}
      \resizebox{0.45\linewidth}{!}{
%
%
\definecolor{mycolor1}{rgb}{0.00000,0.44700,0.74100}%
\definecolor{mycolor2}{rgb}{0.85000,0.32500,0.09800}%
\definecolor{mycolor3}{rgb}{0.92900,0.69400,0.12500}%
\definecolor{mycolor4}{rgb}{0.49400,0.18400,0.55600}%
\begin{tikzpicture}

\begin{axis}[%
width=4.5 cm,
height=4.5 cm,
at={(0 cm,0in)},
scale only axis,
xmin=0,
xmax=11,
xlabel style={font=\color{white!15!black}},
xlabel={$t$ [step]},
ylabel style={font=\color{white!15!black}},
ylabel={$\Gamma(t)$},
ymin=0,
ymax=1,
axis background/.style={fill=white},
xmajorgrids,
ymajorgrids,
legend style={at={(0,1)}, anchor=north west, legend cell align=left, align=left, draw=white!15!black}
]

\addplot[area legend, draw=none, fill=mycolor1, fill opacity=0.15, forget plot]
table[row sep=crcr] {%
x	y\\
0	0.0267857142857143\\
1	0.0287228045398246\\
2	0.034823585344452\\
3	0.0562376013585175\\
4	0.138474481322697\\
5	0.197449325259987\\
6	0.274156767247821\\
7	0.370536358485805\\
8	0.449518085386686\\
9	0.527178845920555\\
10	0.596882111340693\\
11	0.665381427737722\\
11	0.773904286547992\\
10	0.724546460087879\\
9	0.676392582650874\\
8	0.598696200327599\\
7	0.508035070085623\\
6	0.390128947037893\\
5	0.281122103311442\\
4	0.188311232963017\\
3	0.127690970070054\\
2	0.0758907003698337\\
1	0.037348624031604\\
0	0.0267857142857143\\
}--cycle;
\addplot [color=mycolor1, line width=1.5pt]
  table[row sep=crcr]{%
0	0.0267857142857143\\
1	0.0330357142857143\\
2	0.0553571428571429\\
3	0.0919642857142857\\
4	0.163392857142857\\
5	0.239285714285714\\
6	0.332142857142857\\
7	0.439285714285714\\
8	0.524107142857143\\
9	0.601785714285714\\
10	0.660714285714286\\
11	0.719642857142857\\
};
\addlegendentry{CRD}

\addplot[area legend, draw=none, fill=mycolor2, fill opacity=0.15, forget plot]
table[row sep=crcr] {%
x	y\\
0	0.0267857142857143\\
1	0.037774147163893\\
2	0.114541310714787\\
3	0.206658616063121\\
4	0.314117565951679\\
5	0.420531205078727\\
6	0.535049420873476\\
7	0.618866930972409\\
8	0.690531115680057\\
9	0.779961605868045\\
10	0.874918079612023\\
11	0.910615210244396\\
11	0.985813361184176\\
10	0.962581920387977\\
9	0.902181251274813\\
8	0.813040312891371\\
7	0.729347354741877\\
6	0.63816486484081\\
5	0.538397366349844\\
4	0.441239576905464\\
3	0.295127098222593\\
2	0.169387260713784\\
1	0.0604401385503927\\
0	0.0267857142857143\\
}--cycle;
\addplot [color=mycolor2, line width=1.5pt]
  table[row sep=crcr]{%
0	0.0267857142857143\\
1	0.0491071428571428\\
2	0.141964285714286\\
3	0.250892857142857\\
4	0.377678571428571\\
5	0.479464285714286\\
6	0.586607142857143\\
7	0.674107142857143\\
8	0.751785714285714\\
9	0.841071428571429\\
10	0.91875\\
11	0.948214285714286\\
};
\addlegendentry{RD}

\addplot[area legend, draw=none, fill=mycolor3, fill opacity=0.15, forget plot]
table[row sep=crcr] {%
x	y\\
0	0.0267857142857143\\
1	0.0287228045398246\\
2	0.0586439579137292\\
3	0.180291056244476\\
4	0.3378757157887\\
5	0.521055234621096\\
6	0.681471682734523\\
7	0.834362195151915\\
8	0.961162173342868\\
9	0.983349008356541\\
10	1\\
11	1\\
11	1\\
10	1\\
9	1.00593670592917\\
8	0.976337826657132\\
7	0.970994947705228\\
6	0.793528317265477\\
5	0.584301908236046\\
4	0.435338569925586\\
3	0.266137515184095\\
2	0.119927470657699\\
1	0.037348624031604\\
0	0.0267857142857143\\
}--cycle;
\addplot [color=mycolor3, line width=1.5pt]
  table[row sep=crcr]{%
0	0.0267857142857143\\
1	0.0330357142857143\\
2	0.0892857142857143\\
3	0.223214285714286\\
4	0.386607142857143\\
5	0.552678571428571\\
6	0.7375\\
7	0.902678571428571\\
8	0.96875\\
9	0.994642857142857\\
10	1\\
11	1\\
};
\addlegendentry{CD}

\addplot[area legend, draw=none, fill=mycolor4, fill opacity=0.15, forget plot]
table[row sep=crcr] {%
x	y\\
0	0.0267857142857143\\
1	0.037774147163893\\
2	0.176035634508999\\
3	0.348794810380738\\
4	0.515917870129946\\
5	0.753897572143383\\
6	0.94444966945218\\
7	0.973214285714286\\
8	1\\
9	1\\
10	1\\
11	1\\
11	1\\
10	1\\
9	1\\
8	1\\
7	0.973214285714286\\
6	0.951978901976391\\
5	0.781816713570903\\
4	0.585867844155768\\
3	0.402990903904976\\
2	0.231107222633858\\
1	0.0604401385503927\\
0	0.0267857142857143\\
}--cycle;
\addplot [color=mycolor4, line width=1.5pt]
  table[row sep=crcr]{%
0	0.0267857142857143\\
1	0.0491071428571428\\
2	0.203571428571429\\
3	0.375892857142857\\
4	0.550892857142857\\
5	0.767857142857143\\
6	0.948214285714286\\
7	0.973214285714286\\
8	1\\
9	1\\
10	1\\
11	1\\
};
\addlegendentry{ND}

\end{axis}

\end{tikzpicture}%
} &  \resizebox{0.45\linewidth}{!}{
%
%
\definecolor{mycolor1}{rgb}{0.00000,0.44700,0.74100}%
\definecolor{mycolor2}{rgb}{0.85000,0.32500,0.09800}%
\definecolor{mycolor3}{rgb}{0.92900,0.69400,0.12500}%
\definecolor{mycolor4}{rgb}{0.49400,0.18400,0.55600}%
\begin{tikzpicture}

\begin{axis}[%
width=4.5 cm,
height=4.5 cm,
at={(5.5 cm,0in)},
scale only axis,
xmin=0,
xmax=10,
xlabel style={font=\color{white!15!black}},
xlabel={$t$ [step]},
ylabel style={font=\color{white!15!black}},
ylabel={$\bar{U}(t)$},
ymin=0,
ymax=1,
axis background/.style={fill=white},
xmajorgrids,
ymajorgrids,
legend style={at={(rel axis cs:1,1)}, anchor=north east, legend cell align=left, align=left, draw=white!15!black}
]
\addplot[area legend, draw=none, fill=mycolor1, fill opacity=0.15, forget plot]
table[row sep=crcr] {%
x	y\\
0	0.446428575739033\\
1	0.347255546580087\\
2	0.253490167566966\\
3	0.196337724243942\\
4	0.157178584751309\\
5	0.128887073691809\\
6	0.105094411303102\\
7	0.078438648566158\\
8	0.0650439225974883\\
9	0.0547267553609572\\
10	0.0429236856753908\\
10	0.0615857484027864\\
9	0.0713919922139273\\
8	0.0844249001304039\\
7	0.101945497076995\\
6	0.116228890908088\\
5	0.138293033973591\\
4	0.163403694345397\\
3	0.19827513114801\\
2	0.253773893830267\\
1	0.347255546580087\\
0	0.446428575739033\\
}--cycle;
\addplot [color=mycolor1, line width=1.5pt]
  table[row sep=crcr]{%
0	0.446428575739033\\
1	0.347255546580087\\
2	0.253632030698617\\
3	0.197306427695976\\
4	0.160291139548353\\
5	0.1335900538327\\
6	0.110661651105595\\
7	0.0901920728215763\\
8	0.0747344113639461\\
9	0.0630593737874423\\
10	0.0522547170390886\\
};
\addlegendentry{CRD}

\addplot[area legend, draw=none, fill=mycolor2, fill opacity=0.15, forget plot]
table[row sep=crcr] {%
x	y\\
0	0.446428575739033\\
1	0.347255546580087\\
2	0.243893577685076\\
3	0.172197194994407\\
4	0.120339172692405\\
5	0.081426139811111\\
6	0.0552966752174524\\
7	0.04463581055779\\
8	0.0324688867701721\\
9	0.0192236587520598\\
10	0.00786534656628631\\
10	0.027704171927843\\
9	0.0395799407231314\\
8	0.0441517840214853\\
7	0.0512753187616634\\
6	0.0693751589270559\\
5	0.0995580114542772\\
4	0.137220410944587\\
3	0.184915492691761\\
2	0.253919973850626\\
1	0.347255546580087\\
0	0.446428575739033\\
}--cycle;
\addplot [color=mycolor2, line width=1.5pt]
  table[row sep=crcr]{%
0	0.446428575739033\\
1	0.347255546580087\\
2	0.248906775767851\\
3	0.178556343843084\\
4	0.128779791818496\\
5	0.0904920756326941\\
6	0.0623359170722541\\
7	0.0479555646597267\\
8	0.0383103353958287\\
9	0.0294017997375956\\
10	0.0177847592470646\\
};
\addlegendentry{RD}

\addplot[area legend, draw=none, fill=mycolor3, fill opacity=0.15, forget plot]
table[row sep=crcr] {%
x	y\\
0	0.361039030082181\\
1	0.137904545906874\\
2	0.0525385691268533\\
3	0.0186672016411903\\
4	0.00596115037118378\\
5	0.00175217308079973\\
6	0.000503345813621984\\
7	9.14794661507872e-05\\
8	7.13432383793147e-06\\
9	1.32852449877754e-06\\
10	-1.45630383179754e-07\\
10	4.08457201983568e-07\\
9	2.6816511972066e-06\\
8	2.87815282820535e-05\\
7	0.000145455482667078\\
6	0.000571182147190596\\
5	0.00205920080831843\\
4	0.00665287240602156\\
3	0.0196584519604881\\
2	0.0527104878630921\\
1	0.137904545906874\\
0	0.361039030082181\\
}--cycle;
\addplot [color=mycolor3, line width=1.5pt]
  table[row sep=crcr]{%
0	0.361039030082181\\
1	0.137904545906874\\
2	0.0526245284949727\\
3	0.0191628268008392\\
4	0.00630701138860267\\
5	0.00190568694455908\\
6	0.00053726398040629\\
7	0.000118467474408933\\
8	1.79579260599925e-05\\
9	2.00508784799207e-06\\
10	1.31413409401907e-07\\
};
\addlegendentry{CD}

\addplot[area legend, draw=none, fill=mycolor4, fill opacity=0.15, forget plot]
table[row sep=crcr] {%
x	y\\
0	0.361039030082181\\
1	0.137904545906874\\
2	0.0515675079972442\\
3	0.0163033569498579\\
4	0.00485574226573807\\
5	0.00130941906419642\\
6	0.00025852057942741\\
7	2.57411646725308e-05\\
8	6.02440464938705e-06\\
9	0\\
10	0\\
10	0\\
9	0\\
8	6.02499683025557e-06\\
7	2.94871416257522e-05\\
6	0.00029599918368139\\
5	0.00150909456365058\\
4	0.00528892174420078\\
3	0.0175615661895251\\
2	0.052578983911079\\
1	0.137904545906874\\
0	0.361039030082181\\
}--cycle;
\addplot [color=mycolor4, line width=1.5pt]
  table[row sep=crcr]{%
0	0.361039030082181\\
1	0.137904545906874\\
2	0.0520732459541616\\
3	0.0169324615696915\\
4	0.00507233200496942\\
5	0.0014092568139235\\
6	0.0002772598815544\\
7	2.76141531491415e-05\\
8	6.02470073982131e-06\\
9	0\\
10	0\\
};
\addlegendentry{ND}

\end{axis}

\end{tikzpicture}%
} 
 \end{tabular}      
    }\\   
    \subfigure[Fair policies\label{fig:equi_equa}]{
     \begin{tabular}{cc}
      \resizebox{0.45\linewidth}{!}{
%
%
\definecolor{mycolor1}{rgb}{0.00000,0.44700,0.74100}%
\definecolor{mycolor2}{rgb}{0.85000,0.32500,0.09800}%
\definecolor{mycolor3}{rgb}{0.92900,0.69400,0.12500}%
\definecolor{mycolor4}{rgb}{0.49400,0.18400,0.55600}%
\begin{tikzpicture}

\begin{axis}[%
width=4.5 cm,
height=4.5 cm,
at={(0 cm,0in)},
scale only axis,
xmin=0,
xmax=11,
xlabel style={font=\color{white!15!black}},
xlabel={$t$ [step]},
ylabel style={font=\color{white!15!black}},
ylabel={$\Gamma(t)$},
ymin=0,
ymax=1,
axis background/.style={fill=white},
xmajorgrids,
ymajorgrids,
legend style={at={(0,1)}, anchor=north west, legend cell align=left, align=left, draw=white!15!black}
]

\addplot[area legend, draw=none, fill=mycolor1, fill opacity=0.15, forget plot]
table[row sep=crcr] {%
x	y\\
0	0.0267857142857143\\
1	0.0287228045398246\\
2	0.0288054661537744\\
3	0.0387000793037742\\
4	0.0514618483497826\\
5	0.0699796637484966\\
6	0.0821853982796374\\
7	0.101494660896655\\
8	0.122290058218016\\
9	0.149403721291443\\
10	0.172009968648556\\
11	0.196968566711941\\
11	0.301245719002344\\
10	0.274418602780016\\
9	0.234524850137128\\
8	0.209852798924841\\
7	0.182433910531917\\
6	0.157100316006077\\
5	0.126448907680075\\
4	0.102109580221646\\
3	0.0755856349819401\\
2	0.0497659624176542\\
1	0.037348624031604\\
0	0.0267857142857143\\
}--cycle;
\addplot [color=mycolor1, line width=1.5pt]
  table[row sep=crcr]{%
0	0.0267857142857143\\
1	0.0330357142857143\\
2	0.0392857142857143\\
3	0.0571428571428571\\
4	0.0767857142857143\\
5	0.0982142857142857\\
6	0.119642857142857\\
7	0.141964285714286\\
8	0.166071428571429\\
9	0.191964285714286\\
10	0.223214285714286\\
11	0.249107142857143\\
};
\addlegendentry{CRD}

\addplot[area legend, draw=none, fill=mycolor2, fill opacity=0.15, forget plot]
table[row sep=crcr] {%
x	y\\
0	0.0267857142857143\\
1	0.037774147163893\\
2	0.0847002697816522\\
3	0.144836819884254\\
4	0.198824919614624\\
5	0.286799614475999\\
6	0.367133651199649\\
7	0.434492205832082\\
8	0.496006546589425\\
9	0.541784789051362\\
10	0.578690555661552\\
11	0.623076386279027\\
11	0.803709328006687\\
10	0.757023730052733\\
9	0.67964378237721\\
8	0.605779167696289\\
7	0.549436365596489\\
6	0.506080634514637\\
5	0.42927181409543\\
4	0.336889366099662\\
3	0.235520322972888\\
2	0.136728301646919\\
1	0.0604401385503927\\
0	0.0267857142857143\\
}--cycle;
\addplot [color=mycolor2, line width=1.5pt]
  table[row sep=crcr]{%
0	0.0267857142857143\\
1	0.0491071428571428\\
2	0.110714285714286\\
3	0.190178571428571\\
4	0.267857142857143\\
5	0.358035714285714\\
6	0.436607142857143\\
7	0.491964285714286\\
8	0.550892857142857\\
9	0.610714285714286\\
10	0.667857142857143\\
11	0.713392857142857\\
};
\addlegendentry{RD}

\addplot[area legend, draw=none, fill=mycolor3, fill opacity=0.15, forget plot]
table[row sep=crcr] {%
x	y\\
0	0.0267857142857143\\
1	0.0287228045398246\\
2	0.0586439579137292\\
3	0.180291056244476\\
4	0.3378757157887\\
5	0.521055234621096\\
6	0.681471682734523\\
7	0.834362195151915\\
8	0.961162173342868\\
9	0.983349008356541\\
10	1\\
11	1\\
11	1\\
10	1\\
9	1.00593670592917\\
8	0.976337826657132\\
7	0.970994947705228\\
6	0.793528317265477\\
5	0.584301908236046\\
4	0.435338569925586\\
3	0.266137515184095\\
2	0.119927470657699\\
1	0.037348624031604\\
0	0.0267857142857143\\
}--cycle;
\addplot [color=mycolor3, line width=1.5pt]
  table[row sep=crcr]{%
0	0.0267857142857143\\
1	0.0330357142857143\\
2	0.0892857142857143\\
3	0.223214285714286\\
4	0.386607142857143\\
5	0.552678571428571\\
6	0.7375\\
7	0.902678571428571\\
8	0.96875\\
9	0.994642857142857\\
10	1\\
11	1\\
};
\addlegendentry{CD}

\addplot[area legend, draw=none, fill=mycolor4, fill opacity=0.15, forget plot]
table[row sep=crcr] {%
x	y\\
0	0.0267857142857143\\
1	0.037774147163893\\
2	0.176035634508999\\
3	0.348794810380738\\
4	0.515917870129946\\
5	0.753897572143383\\
6	0.94444966945218\\
7	0.973214285714286\\
8	1\\
9	1\\
10	1\\
11	1\\
11	1\\
10	1\\
9	1\\
8	1\\
7	0.973214285714286\\
6	0.951978901976391\\
5	0.781816713570903\\
4	0.585867844155768\\
3	0.402990903904976\\
2	0.231107222633858\\
1	0.0604401385503927\\
0	0.0267857142857143\\
}--cycle;
\addplot [color=mycolor4, line width=1.5pt]
  table[row sep=crcr]{%
0	0.0267857142857143\\
1	0.0491071428571428\\
2	0.203571428571429\\
3	0.375892857142857\\
4	0.550892857142857\\
5	0.767857142857143\\
6	0.948214285714286\\
7	0.973214285714286\\
8	1\\
9	1\\
10	1\\
11	1\\
};
\addlegendentry{ND}

\end{axis}
\end{tikzpicture}} &  \resizebox{0.45\linewidth}{!}{
%
%
\definecolor{mycolor1}{rgb}{0.00000,0.44700,0.74100}%
\definecolor{mycolor2}{rgb}{0.85000,0.32500,0.09800}%
\definecolor{mycolor3}{rgb}{0.92900,0.69400,0.12500}%
\definecolor{mycolor4}{rgb}{0.49400,0.18400,0.55600}%
\begin{tikzpicture}

\begin{axis}[%
width=4.5 cm,
height=4.5 cm,
at={(5.5 cm,0in)},
scale only axis,
xmin=0,
xmax=10,
xlabel style={font=\color{white!15!black}},
xlabel={$t$ [step]},
ylabel style={font=\color{white!15!black}},
ylabel={$\bar{U}(t)$},
ymin=0,
ymax=1,
axis background/.style={fill=white},
xmajorgrids,
ymajorgrids,
legend style={at={(rel axis cs:1,1)}, anchor=north east, legend cell align=left, align=left, draw=white!15!black}
]

\addplot[area legend, draw=none, fill=mycolor1, fill opacity=0.15, forget plot]
table[row sep=crcr] {%
x	y\\
0	0.257996324953038\\
1	0.15827527083987\\
2	0.102540763344057\\
3	0.0794014299788035\\
4	0.0624341200765631\\
5	0.0534630413388889\\
6	0.0461686401762571\\
7	0.0395216471171678\\
8	0.0352037030147938\\
9	0.0310303654543664\\
10	0.0278616168489915\\
10	0.0320730803097198\\
9	0.0379975570352353\\
8	0.0413331793760907\\
7	0.0458226948382035\\
6	0.0509642147720627\\
5	0.0582351648103182\\
4	0.0665757458815117\\
3	0.0854476896898813\\
2	0.110850988108007\\
1	0.15827527083987\\
0	0.257996324953038\\
}--cycle;
\addplot [color=mycolor1, line width=1.5pt]
  table[row sep=crcr]{%
0	0.257996324953038\\
1	0.15827527083987\\
2	0.106695875726032\\
3	0.0824245598343424\\
4	0.0645049329790374\\
5	0.0558491030746035\\
6	0.0485664274741599\\
7	0.0426721709776857\\
8	0.0382684411954422\\
9	0.0345139612448008\\
10	0.0299673485793557\\
};
\addlegendentry{CRD}

\addplot[area legend, draw=none, fill=mycolor2, fill opacity=0.15, forget plot]
table[row sep=crcr] {%
x	y\\
0	0.257996324953038\\
1	0.15827527083987\\
2	0.100617568268031\\
3	0.0743787244959927\\
4	0.0489574904278205\\
5	0.0375664545970635\\
6	0.0290643982595159\\
7	0.0194615480546419\\
8	0.0134913985487521\\
9	0.00962992785329211\\
10	0.00802054064660915\\
10	0.0173139693242423\\
9	0.0197356594494783\\
8	0.0262462460600806\\
7	0.0292375015580494\\
6	0.0436771636335952\\
5	0.0506330496044664\\
4	0.0596829322932648\\
3	0.0820411107016033\\
2	0.103682714470948\\
1	0.15827527083987\\
0	0.257996324953038\\
}--cycle;
\addplot [color=mycolor2, line width=1.5pt]
  table[row sep=crcr]{%
0	0.257996324953038\\
1	0.15827527083987\\
2	0.10215014136949\\
3	0.078209917598798\\
4	0.0543202113605426\\
5	0.0440997521007649\\
6	0.0363707809465556\\
7	0.0243495248063456\\
8	0.0198688223044164\\
9	0.0146827936513852\\
10	0.0126672549854257\\
};
\addlegendentry{RD}

\addplot[area legend, draw=none, fill=mycolor3, fill opacity=0.15, forget plot]
table[row sep=crcr] {%
x	y\\
0	0.361039024346516\\
1	0.137904581625233\\
2	0.0525386609695348\\
3	0.0186671212398886\\
4	0.00596113750112128\\
5	0.00175227724049991\\
6	0.000501814428766538\\
7	9.16047738051249e-05\\
8	7.89283028068885e-06\\
9	1.28773238458722e-06\\
10	-1.71550479934446e-07\\
10	4.81156417078037e-07\\
9	3.05547037564855e-06\\
8	2.71017919281739e-05\\
7	0.000145618735770564\\
6	0.00056950840499493\\
5	0.00205933847862732\\
4	0.00665296836416386\\
3	0.0196582724969611\\
2	0.0527106293570242\\
1	0.137904581625233\\
0	0.361039024346516\\
}--cycle;
\addplot [color=mycolor3, line width=1.5pt]
  table[row sep=crcr]{%
0	0.361039024346516\\
1	0.137904581625233\\
2	0.0526246451632795\\
3	0.0191626968684249\\
4	0.00630705293264257\\
5	0.00190580785956361\\
6	0.000535661416880734\\
7	0.000118611754787844\\
8	1.74973111044314e-05\\
9	2.17160138011789e-06\\
10	1.54802968571795e-07\\
};
\addlegendentry{CD}

\addplot[area legend, draw=none, fill=mycolor4, fill opacity=0.15, forget plot]
table[row sep=crcr] {%
x	y\\
0	0.361039024346516\\
1	0.137904581625233\\
2	0.051567702917243\\
3	0.0163033851521248\\
4	0.00485565608867875\\
5	0.00130942524617644\\
6	0.00025763967842791\\
7	2.58393214502164e-05\\
8	5.752149348566e-06\\
9	0\\
10	0\\
10	0\\
9	0\\
8	5.76053160837518e-06\\
7	2.95748827196919e-05\\
6	0.000295093920663027\\
5	0.00150919572490944\\
4	0.00528890033297166\\
3	0.0175615138714101\\
2	0.0525791865917892\\
1	0.137904581625233\\
0	0.361039024346516\\
}--cycle;
\addplot [color=mycolor4, line width=1.5pt]
  table[row sep=crcr]{%
0	0.361039024346516\\
1	0.137904581625233\\
2	0.0520734447545161\\
3	0.0169324495117675\\
4	0.00507227821082521\\
5	0.00140931048554294\\
6	0.000276366799545469\\
7	2.77071020849542e-05\\
8	5.75634047847059e-06\\
9	0\\
10	0\\
};
\addlegendentry{ND}

\end{axis}

\end{tikzpicture}%
} 
    \end{tabular}
    } 
\end{tabular}\vspace{-.2cm}
\caption{Adoption rate \emph{vs} average policy over $N_{\text{MC}}$ Monte Carlo simulations for the different deficit scenarios (ND, RD, CD, and CRD). On the left panel, we always report the mean values (solid lines) and standard deviations (shaded areas) of the adoption rate over time.}
\label{fig:fair_vs_nofair}
\end{figure}

The results obtained for all aforementioned scenarios are summarized in \figurename{~\ref{fig:fair_vs_nofair}}. The latter shows the mean values (solid lines) and standard deviations (shaded areas) of the acceptance rates for the new technology 
\begin{equation}
    \Gamma(t)=\frac{1}{N}\sum_{v \in \mathcal{V}}x_v(t) \in [0,1],
\end{equation}
over $t \in [0,T]$ and across our $N_{\text{MC}}$ Monte Carlo simulations, as well as the average nudging policy
\begin{equation}
    \bar{U}(t)=\frac{1}{N_{\text{MC}}}\sum_{i=1}^{N_{\text{MC}}}\left[\frac{1}{N}\sum_{v \in \mathcal{V}} u_{v}^{i}(t)\right],
\end{equation}
with $u_{v}^{i}(t) \in [0,1]$ indicating the nudging policy enacted on the $v$-th agent in $t \in [0,T-1]$ over the $i$-th Monte Carlo simulation, with $v \in \mathcal{V}$ and $i \in [1,N_{\text{MC}}]$. Looking at the achieved adoption rates in the two cases, the CRD is the most challenging scenario for nudging adoption, as the population is generally less receptive to policies and their adoption is eventually guided by social prejudices. Meanwhile, the one-sided and fair policies yield nearly identical adoption rates over time in the ND and CD scenarios. This outcome occurs because the MPC can predict agents' reluctance by leveraging its knowledge of their receptivity to policies, and compensates for it by increasing the value of $u_v(t)$ to less receptive agents $v \in \mathcal{V}$. By contrast, credibility deficits influence the diffusion process through social perception mechanisms that are not directly mitigable by the policies. Higher adoption levels are then achieved in both the RD and CRD scenarios when unfairness is not penalized. This result comes at the price of a higher policy effort (see \figurename{~\ref{fig:no_fair}}), indicating that our design choices in \tablename{~\ref{tab:parameters}} yield a policy design scheme that prioritizes performance when fairness is not a factor. Meanwhile, in our simulations, a receptivity deficit seems to reduce the adoption rate more than a credibility one when weighing equality and equity in policy design (see \figurename{~\ref{fig:equi_equa}). This outcome is particularly evident at the end of the simulation horizon, as expected, since fairness (and especially equality) tends to reduce the average input to the agents, and the receptivity deficit tends to reduce the impact of policies on the evolution of individual reluctances.  

\begin{figure}[!tb]
\centering
\begin{tabular}{cc}  

    \subfigure[One-sided policies]{%
        \label{fig:fairness_maps_nofair}
        \includegraphics[width=0.45\linewidth]{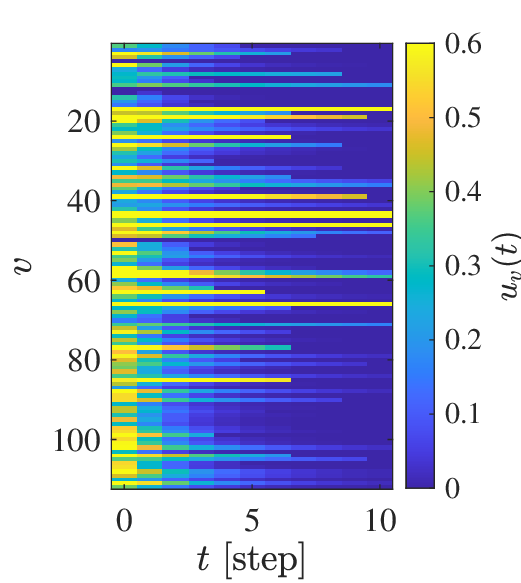}
    } &
    \subfigure[Fair policies]{%
        \label{fig:fairness_maps_fair}
        \includegraphics[width=0.45\linewidth]{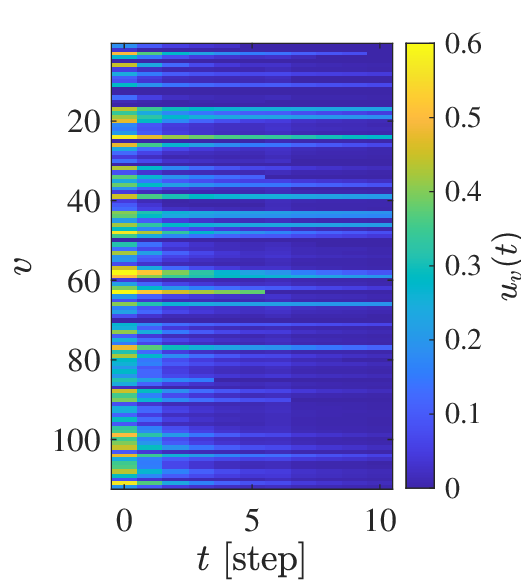}
    } \\[0.2cm]

    \subfigure[Equity only policy ($N=0$)]{%
        \label{fig:fairness_maps_equi}
        \includegraphics[width=0.45\linewidth]{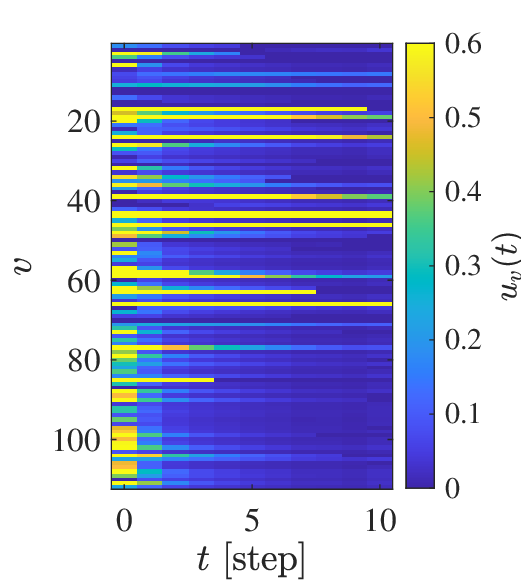}
    }    &
    
    \subfigure[Equality only policy ($M\!\!=\!\!0$)]{%
        \label{fig:fairness_maps_equa}
        \includegraphics[width=0.45\linewidth]{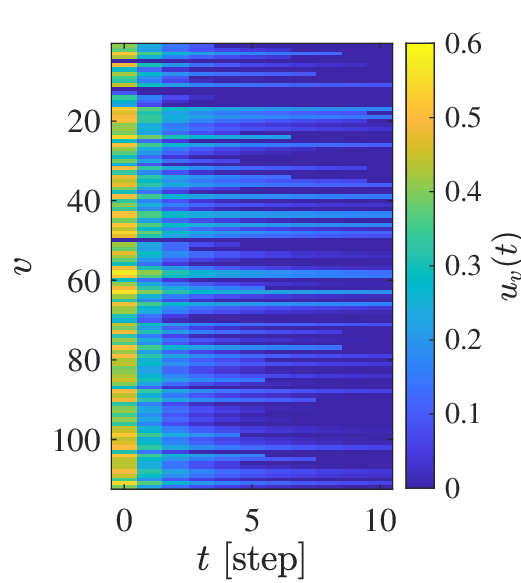}
    } 

\end{tabular}
\caption{Nudging policy received by each agent at each time instant over one of the $N_{\mathrm{MC}}=10$ Monte Carlo simulations. Each row corresponds to an agent, each column to a time step, while the color itself indicates the policy magnitude. The lower the nudging policy, the more the color turns towards blue.}
\label{fig:fairness_maps}
\end{figure}



    


To further understand the impact of fairness on resource allocation, we compare their distribution over one of the $N_{\text{MC}}$ Monte Carlo simulations in the presence of both credibility and receptivity deficits across 4 scenarios: the one-sided and fair cases, as well as those in which either equality or equity is accounted for in policy design. As shown in \figurename{~\ref{fig:fairness_maps}}, policy efforts are visibly different across agents when equality is not accounted for (see \figurename{~\ref{fig:fairness_maps_nofair}} and \figurename{~\ref{fig:fairness_maps_equi}}). Therefore, solving \eqref{eq:MPC} without fairness yields a policy that focuses on reducing heterogeneity in reluctance, resulting in stronger, more targeted early interventions for disadvantaged agents. A feature that is preserved when equity is also considered, as it is clear from \figurename{~\ref{fig:fairness_maps_equi}}. Differences in policy allocation are mitigated when equality is also considered, albeit at the cost of a less widespread adoption (as already shown in \figurename{~\ref{fig:fair_vs_nofair}}). Indeed, when only equality is considered, or it is combined with equality (\figurename{~\ref{fig:fairness_maps_equa}} and \figurename{~\ref{fig:fairness_maps_fair}), the controller still compensates for disadvantaged agents, but does so while maintaining an overall balanced distribution of incentives. Note that, as more agents adopt, the dominant color in \figurename{~\ref{fig:fairness_maps}} becomes blue, spotlighting one of the features of the resource allocation strategy discussed in Section~\ref{sec:control}.

\section{Conclusions} \label{sec:conclusions}
In this work, we propose a stochastic innovation diffusion model that formalizes how adoption processes can be affected by structural inequalities and inherent prejudices. Based on this model, we present a strategy for designing nudging policies to overcome adoption barriers via MPC, which penalizes a fairness cost to favor equity in adoption and equality in resource distribution in the face of population heterogeneity. Through simulations based on real mobility-habit data, we show the impact of individual features, social prejudice, as well as a fairness-aware policy design on the innovation diffusion process. In particular and as expected, we show how equality smooths the distribution of nudging interventions across individuals, while equity shapes \emph{to whom} policies are delivered. 

Future work will focus on evaluating alternative fairness metrics for policy design, exploring alternative policy design strategies and their cost-benefit trade-offs, and testing our solutions in other scenarios of interest.

\bibliography{main}             
       
\end{document}